\documentclass[aps,pra,reprint, amsmath, amssymb,superscriptaddress,nofootinbib]{revtex4-1}

\usepackage{bm}
\usepackage[retainorgcmds]{IEEEtrantools}
\usepackage{graphicx}
\usepackage{mathrsfs}
\usepackage{amsmath}
\usepackage{amssymb}
\usepackage{color}
\usepackage{amsfonts}
\usepackage{times,txfonts}
\usepackage{nicefrac}
\usepackage[colorlinks=true,linkcolor=blue,urlcolor=blue,citecolor=blue,pdfusetitle]{hyperref}
\usepackage{amsmath}

\newcommand{\tr}{\text{tr}}

\begin{document}

\title{Quantum relative entropy uncertainty relation}
\date{\today}
\author{Domingos S. P. Salazar}
\affiliation{Unidade de Educa\c c\~ao a Dist\^ancia e Tecnologia,
Universidade Federal Rural de Pernambuco,
52171-900 Recife, Pernambuco, Brazil}

\begin{abstract}
For classic systems, the thermodynamic uncertainty relation (TUR) states that the fluctuations of a current have a lower bound in terms of the entropy production. Some TURs are rooted in information theory, particularly derived from relations between observations (mean and variance) and dissimilarities, such as the Kullback-Leibler divergence, which plays the role of entropy production in stochastic thermodynamics. We generalize this idea for quantum systems, where we find a lower bound for the uncertainty of quantum observables given in terms of the quantum relative entropy. We apply the result to obtain a quantum thermodynamic uncertainty relation in terms of the quantum entropy production, valid for arbitrary dynamics and non-thermal environments.
\end{abstract}
\maketitle{}


{\bf \emph{Introduction -}}
Entropy production is the main concept of thermodynamics far from equilibrium. This concept has been defined and explored extensively in stochastic thermodynamics, where entropy production $\Sigma$ and physical observables become random variables at trajectory level \cite{RevModPhys.93.035008,Campisi2011,Seifert2012Review,Campisi2011,Esposito2009,Ciliberto2013,Crooks1998,Crooks1999,Hanggi2015,Batalhao2014}. In this case, the second law of thermodynamics is stated as
\begin{equation}
\langle \Sigma \rangle \geq 0.
\end{equation}
Among the cornerstones of stochastic thermodynamics, there are the thermodynamic uncertainty relations (TURs) \cite{Barato2015A,Gingrich2016,Polettini2017,Pietzonka2017,Hasegawa2019,Hasegawa2019b,VanTuan2020,Van_Vu_2020,
Timpanaro2019b,Liu2020,Horowitz2022,Potts2019,Proesmans2019,Gianluca2022,Salazar2022d}, which usually take the form
\begin{equation}
\label{TUR}
\frac{\langle \phi^2 \rangle - \langle \phi \rangle^2}{\langle \phi \rangle^2} \geq f(\langle \Sigma \rangle),
\end{equation}
for a current $\phi$, where $f$ is a known function. The TUR establishes that there's always an inherent minimum fluctuation (or uncertainty) in a process that isn't reversible, $\langle \Sigma \rangle \geq 0$. This uncertainty is quantified as the ratio of the variance to the mean squared, observed in the lhs of (\ref{TUR}), and the bound is given solely as function of the entropy production. 

In recent years, there have been significant advancements in extending TURs to the quantum realm. These quantum TURs establish connections between fluctuations and irreversibility, expanding our understanding beyond classical contexts \cite{Brandner2018,Carollo2019,Liu2019,VanVu2022b,Miller2021}, for steady states \cite{Guarnieri2019,Hasegawa2019a}, for the Lindblad's dynamics \cite{hasegawa2021,Hasegawa2023}, and for general open quantum systems \cite{Hasegawa2021b} usually in terms of quantities other than the quantum entropy production. 

An even more direct generalization of (\ref{TUR}) to quantum thermodynamics would benefit from (i) a bound given in terms of the quantum entropy production itself and (ii) valid at strong coupling, for any dynamics. In this sense, we first obtain our main result: for any density matrices, $\rho$ and $\sigma$, and for any Hermitian operator $\hat{\theta}$, we have
\begin{equation}
\label{qTUR}
\frac{\langle \hat{\theta}^2 \rangle_\rho - \langle \hat{\theta} \rangle_\rho^2 + \langle \hat{\theta}^2 \rangle_\sigma - \langle \hat{\theta} \rangle_\sigma^2}{(1/2)(\langle \hat{\theta} \rangle_\rho - \langle \hat{\theta}\rangle_\sigma)^2} \geq f\big(\frac{S(\rho||\sigma)+S(\sigma||\rho)}{2}\big),
\end{equation}
for $\langle \hat{\theta} \rangle_\rho := \tr(\rho \hat{\theta}) \neq \langle \hat{\theta} \rangle_\sigma := \tr(\sigma \hat{\theta})$, $f(x)=1/\sinh^2(g(x)/2)$ and $g(x)$ is the inverse of $h(x):=x\tanh(x/2)$ for $x>0$. $S(\rho||\sigma)=\tr(\rho(\log \rho - \log \sigma))$ is the quantum relative entropy. The bound (\ref{qTUR}) is saturated by a minimal two-level system with commuting operators $\rho, \sigma, \hat{\theta}$. However, in general, (\ref{qTUR}) is not an identity, as we show in the numeric simulations.

With our main result (\ref{qTUR}) in hands, we now turn to a general setup of quantum thermodynamics \cite{RevModPhys.93.035008}, where system and environment are prepared in arbitrary states $\rho_S$ and $\rho_E$, followed by a unitary evolution, such that the final state is entangled and given by $\rho:=\mathcal{U}(\rho_S \otimes \rho_E)\mathcal{U}^\dagger$. After the evolution, we define the reduced state of the system $\rho_S':=\tr_E(\rho)$. In this notation, the quantum entropy production is defined as \cite{Esposito2010a,Manzano2018PRX},
\begin{equation}
\label{qthermo1}
\Sigma :=S(\mathcal{U}(\rho_S \otimes \rho_E)\mathcal{U}^\dagger||\rho_S' \otimes \rho_E)=S(\rho||\sigma),
\end{equation}
which is a dissimilarity between the final state of the forward process $\rho:=\mathcal{U}(\rho_S \otimes \rho_E)\mathcal{U}^\dagger$ and a specific choice for the initial state of the backward process $\sigma:=\rho_S' \otimes \rho_E$. We now define a dual of the entropy production as the following dissimilarity \begin{equation}
\label{qthermo2}
\Sigma^*:=S(\sigma||\rho)=S(\mathcal{U^\dagger}\sigma\mathcal{U}||\rho_S \otimes \rho_E),
\end{equation}
where the last identity used the fact that $\mathcal{U}$ is unitary. Note that $\Sigma^*$ is uniquely defined from $\Sigma$ and $\Sigma^{**}=\Sigma$. Perhaps not surprisingly, the dual (\ref{qthermo2}) is also given in terms of an average stochastic entropy production, as it happens with $\Sigma$ \cite{Manzano2018PRX}, as discussed later on. However, $\Sigma^*$ is not to be confused with the entropy production of the backward process. As a matter of fact, the specific form of $\Sigma^*$ allows us to apply our main result (\ref{qTUR}) for any quantum observable $\hat{\theta}$ acting on the system + environment, using (\ref{qthermo1}) and (\ref{qthermo2}),
\begin{equation}
\label{qthermo3}
\frac{\langle \hat{\theta}^2 \rangle_\rho - \langle \hat{\theta} \rangle_\rho^2 + \langle \hat{\theta}^2 \rangle_\sigma - \langle \hat{\theta} \rangle_\sigma^2}{(1/2)(\langle \hat{\theta} \rangle_\rho - \langle \hat{\theta}\rangle_\sigma)^2} \geq f\big(\frac{\Sigma+\Sigma^*}{2}\big),
\end{equation}
which is our second main result and highlights the role played by $\Sigma^*$ in thermodynamics. The quantum thermodynamic uncertainty relation expressed in (\ref{qthermo3}) is notably general. It covers a quantum thermodynamics framework that accommodates strong coupling and remains valid even when arbitrarily far from equilibrium. Furthermore, it's defined explicitly in terms of the entropy production and its dual. It also recovers other classic TURs \cite{Timpanaro2019b,Salazar2022d} as limiting cases. 

The paper is organized as follows. First, we present the formalism and prove (\ref{qTUR}), which is a result in quantum information. Then, we test the theoretic result with Monte Carlo simulations with two random qubits and a random observable in the presence of coherence, where the bound is verified. We also discuss the saturation of the bound, the role of coherence between $\rho$ and $\sigma$, followed by applications to arbitrary quantum channels and quantum thermodynamics. 

{\bf \emph{Formalism -}}
The idea behind the proof of (\ref{qTUR}) goes as follows. First, we find a lower bound for the lhs of (\ref{qTUR}) in terms of a classic uncertainty, with probabilities $P,Q$ and a complex random variable $\Theta$. Then, we use a result from information theory, which is a lower bound for such classic uncertainty in terms of the symmetric Kullback-Leibler (KL) divergence of $P$ and $Q$. Finally, we show that, for our specific choices of $P$ and $Q$, the symmetric KL equals the symmetric quantum relative entropy between $\rho$ and $\sigma$ and that ends the proof. Details are given below.

Let $\rho$ and $\sigma$ be any density matrices (Hermitian, semi-positive and $\tr(\rho)=\tr(\sigma)=1$). Let $\hat{\theta}^\dagger=\hat{\theta}$ be any Hermitian operator with $\langle \hat{\theta} \rangle_\rho \neq \langle \hat{\theta}\rangle_\sigma$. We have the spectral decomposition, $\rho = \sum_i p_i |p_i\rangle \langle p_i|$ and $\sigma = \sum_j q_j |q_j\rangle \langle q_j|$, with $0\leq p_i, q_j \leq 1$, $\langle p_i |p_j \rangle=\delta_{ij}$ and $\langle q_i |q_j \rangle=\delta_{ij}$. The expected value of $\hat{\theta}$ with respect to $\rho$ is
\begin{equation}
\label{form1}
\tr(\rho \hat{\theta}) = \sum_i p_i \langle p_i | \hat{\theta}|p_i\rangle = \sum_{ij} p_i \langle p_i | \hat{\theta}|q_j\rangle \langle q_j | p_i \rangle,
\end{equation}
and the expression above can be written as 
\begin{equation}
\label{form2}
    \sum_{ij} p_i \langle p_i | \hat{\theta}|q_j\rangle \langle q_j | p_i \rangle = \sum_{ij; \langle q_j |p_i \rangle \neq 0} p_i |\langle q_j | p_i \rangle|^2 \frac{\langle p_i | \hat{\theta}|q_j\rangle}{\langle p_i |q_j \rangle},
\end{equation}
where we used $\langle p_i | q_j \rangle = \langle q_j | p_i \rangle^*$. Now we define $P_{ij}:=p_i |\langle q_j | p_i \rangle|^2$ for all $(i,j)$ and define $\Theta_{ij}:=\langle p_i | \hat{\theta}|q_j\rangle/\langle p_i |q_j \rangle$, if $\langle p_i |q_j \rangle \neq 0$ and $\Theta_{ij}:=0$, if $\langle p_i |q_j \rangle=0$. In terms of $P$ and $\Theta$, we have from (\ref{form1}) and (\ref{form2}),
\begin{equation}
\label{form3}
\tr(\rho \hat{\theta}) = \sum_{ij}P_{ij}\Theta_{ij} := \langle \Theta \rangle_P,
\end{equation}
where we note that $P$ is a probability function, $0 \leq P_{ij}\leq 1$ and $\sum_{ij} P_{ij} = \sum_{ij}p_i \langle q_j | p_i \rangle \langle p_i | q_j \rangle = \tr(\rho) =1$. Similarly, we obtain for the expected value of $\hat{\theta}$ with respect to $\sigma$,
\begin{equation}
\label{form4}
\tr(\sigma \hat{\theta}) = \sum_{ij}Q_{ij}\Theta_{ij} := \langle \Theta \rangle_Q,
\end{equation}
for $Q_{ij}=q_j |\langle q_j | p_i \rangle|^2$, which is also a probability function, $0 \leq Q_{ij} \leq 1$ and $\sum_{ij} {Q_{ij}}=\tr(\sigma)=1$. Analogously, we have for the expected value of $\hat{\theta}^2$ with respect to $\rho$,
\begin{equation}
\label{form5}
\tr(\rho \hat{\theta}^2) = \sum_{ij} p_i \langle p_i |\hat{\theta}|q_j \rangle \langle q_j| \hat{\theta}|p_i\rangle = \sum_{ij} p_i |\langle p_i |\hat{\theta}|q_j \rangle|^2,
\end{equation}
where we used $\hat{\theta}=\hat{\theta}^\dagger$. Then, note that 
\begin{equation}
\label{form6}
\sum_{ij} p_i |\langle p_i |\hat{\theta}|q_j \rangle|^2 \geq \sum_{ij;\langle q_j|p_i\rangle \neq 0} p_i |\langle p_i |\hat{\theta}|q_j \rangle|^2 = \sum_{ij} P_{ij}|\Theta_{ij}|^2,
\end{equation}
which yields after combining (\ref{form5}) and (\ref{form6}),
\begin{equation}
\label{form7}
\tr(\rho \hat{\theta}^2) \geq \sum_{ij} P_{ij}|\Theta_{ij}|^2 :=\langle |\Theta|^2 \rangle_P.
\end{equation}
We have a similar expression in terms of $\sigma$,
\begin{equation}
\label{form8}
\tr(\sigma \hat{\theta}^2) \geq \sum_{ij} Q_{ij}|\Theta_{ij}|^2 :=\langle |\Theta|^2 \rangle_Q.
\end{equation}
Combining expressions (\ref{form3}), (\ref{form4}), (\ref{form7}) and (\ref{form8}), one obtains
\begin{eqnarray}
\label{form9}
\frac{\langle \hat{\theta}^2 \rangle_\rho - \langle \hat{\theta} \rangle_\rho^2 + \langle \hat{\theta}^2 \rangle_\sigma - \langle \hat{\theta} \rangle_\sigma^2}{(1/2)(\langle \hat{\theta} \rangle_\rho - \langle \hat{\theta}\rangle_\sigma)^2} \geq \nonumber
\\
\frac{\langle |\Theta|^2 \rangle_P - |\langle \Theta \rangle_P|^2 + \langle |\Theta|^2 \rangle_Q - |\langle \Theta \rangle_Q|^2}{(1/2)|\langle \Theta \rangle_P - \langle \Theta \rangle_Q|^2},
\end{eqnarray}
which completes the first part of the proof. 

In the second part of the proof, we import a result from information theory \cite{Salazar2022d,Nishiyama2022} and modify it to include complex random variables. For any probabilities $P,Q$ and complex random variable $\Theta$, with $\langle \Theta \rangle_P \neq \langle \Theta \rangle_Q$, the theorem states that
\begin{eqnarray}
\label{form10}
\frac{\langle |\Theta|^2 \rangle_P - |\langle \Theta \rangle_P|^2 + \langle |\Theta|^2 \rangle_Q - |\langle \Theta \rangle_Q|^2}{(1/2)|\langle \Theta \rangle_P - \langle \Theta \rangle_Q|^2} \geq f(\tilde{D}(P,Q)),
\end{eqnarray}
where $\tilde{D}(P,Q):=(D(P|Q)+D(Q|P))/2$ is the symmetric KL divergence and $D(P|Q)=\sum_s P(s)\log(P(s)/Q(s))$ is the KL divergence, and $f(x)=\sinh(g(x)/2)^{-2}$ and $g(x)$ is the inverse of $h(x)=x\tanh(x/2)$ for $x>0$. The proof of (\ref{form10}) is given in the Appendix.

Finally, for the third part of the proof, take again $P_{ij}=p_i |\langle q_j | p_i \rangle|^2$ and $Q_{ij}=q_j |\langle q_j | p_i \rangle|^2$. In this case, we have
\begin{equation}
\label{form11}
D(P|Q)=\sum_{ij}P_{ij} \log \frac{P_{ij}}{Q_{ij}} = \sum_{ij} |\langle q_j | p_i \rangle|^2 p_i \log\frac{p_i}{q_j},
\end{equation}
and after using $\sum_j|\langle q_j | p_i \rangle|^2=1$, eq. (\ref{form11}) simplifies to
\begin{equation}
\label{form12}
D(P|Q)=\sum_{i} p_i \log p_i -  \sum_{ij}|\langle q_j | p_i \rangle|^2 p_i \log q_j=S(\rho||\sigma).
\end{equation}
Similarly, we have $D(Q|P)=S(\sigma||\rho)$ and the following identity
\begin{equation}
\label{form13}
\tilde{D}(P,Q)=\frac{1}{2}\big(S(\rho||\sigma)+S(\sigma||\rho) \big) := \tilde{S}(\rho,\sigma).
\end{equation}
Combining (\ref{form9}), (\ref{form10}) and (\ref{form13}), we obtain our main result (\ref{qTUR}). The quantum thermodynamics application (\ref{qthermo3}) follows immediately from the definitions of the quantum entropy production (\ref{qthermo1}) and the dual (\ref{qthermo2}), where we used
\begin{equation}
S(\mathcal{U}^\dagger \sigma \mathcal{U}||\rho_S \otimes \rho_E)=S(\sigma||\mathcal{U}(\rho_S \otimes \rho_E)\mathcal{U}^\dagger)=S(\sigma||\rho).
\end{equation}

{\bf \emph{Discussion -}}
Let us discuss the meaning and the broad scope of (\ref{qTUR}). First, we note that the form of the lhs of (\ref{qTUR}) resembles the uncertainty of classic TURs. We define
\begin{equation}
\label{disc0}
U(\hat{\theta}; \rho, \sigma):=\frac{\langle \hat{\theta}^2 \rangle_\rho - \langle \hat{\theta} \rangle_\rho^2 + \langle \hat{\theta}^2 \rangle_\sigma - \langle \hat{\theta} \rangle_\sigma^2}{(1/2)(\langle \hat{\theta} \rangle_\rho - \langle \hat{\theta}\rangle_\sigma)^2}
\end{equation}
as a type of quantum uncertainty of the observable $\hat{\theta}$ with respect to two states $\rho$ and $\sigma$. By definition, this uncertainty is symmetric, $U(\hat{\theta};\rho,\sigma)=U(\hat{\theta};\sigma,\rho)$, as in other quantum uncertainty relations \cite{Hasegawa2023}. Using the notation (\ref{disc0}), relation (\ref{qTUR}) may be presented as a lower bound for the symmetric quantum relative entropy in terms of any observable $\hat{\theta}$, 
\begin{equation}
\label{disc1}
\tilde{S}(\rho,\sigma) \geq B\Big(U(\hat{\theta};\rho, \sigma)\Big),
\end{equation}
where $B(x):=2 (1+x)^{-1/2} \tanh^{-1}[(1+x)^{-1/2}]=(1+x)^{-1/2}\log[(\sqrt{x+1}+1)/(\sqrt{x+1}-1)]$, which might be useful in situations where the statistics of any $\hat{\theta}$ is easier to compute. In the specific case $\langle \hat{\theta} \rangle_\sigma = - \langle \hat{\theta} \rangle_\rho$ and $\langle \hat{\theta}^2 \rangle_\sigma = \langle \hat{\theta}^2 \rangle_\rho$, we get 
\begin{equation}
\label{disc2}
U(\hat{\theta};\rho,\sigma)=\frac{\langle \hat{\theta}^2 \rangle_\rho - \langle \hat{\theta}\rangle_\rho^2}{\langle \hat{\theta}\rangle_\rho^2}\geq f(\tilde{S}(\rho,\sigma)),
\end{equation}
which corresponds to the uncertainty of classic currents in the exchange TUR (\ref{TUR}) \cite{Timpanaro2019b}. More generally, the absence of coherence, $[\rho,\sigma]=0$ and $[\hat{\theta},\rho]=0$, reduces $U(\hat{\theta};\rho,\sigma)$ in (\ref{disc1}) to the uncertainty used in other classic generalizations of the exchange TUR, such as the hysteretic TUR \cite{Potts2019,Proesmans2019,Gianluca2022, Salazar2022d}.

The analogy with classic TURs immediately suggests which quantum system would saturate the bound, 
$U(\hat{\theta};\rho,\sigma)=f(\tilde{S}(\rho,\sigma))$. As in the classic case, the bound in (\ref{qTUR}) is saturated for a specific minimal two-level system. Consider $\rho=2\cosh(\epsilon/2)[e^{\epsilon/2}|1\rangle \langle 1| + e^{-\epsilon/2}|0\rangle \langle 0|]$,  $\sigma=2\cosh(\epsilon/2)[e^{-\epsilon/2}|1\rangle \langle 1| + e^{\epsilon/2}|0\rangle \langle 0|]$ and $\hat{\theta}=\omega(|1\rangle \langle 1| - |0\rangle \langle 0|)$. In this case, one has $\tr(\rho \hat{\theta})=\omega \tanh(\epsilon/2)$,  $\tr(\sigma\hat{\theta})=-\omega \tanh(\epsilon/2)$ and $\tr(\rho \hat{\theta}^2)=\tr(\sigma \hat{\theta}^2)=\omega^2$, such that
\begin{equation}
\label{disc4}
U(\hat{\theta};\rho, \sigma) = \sinh^{-2}(\epsilon/2)=f(h(\epsilon))=f(\tilde{S}(\rho,\sigma)),
\end{equation}
since $\tilde{S}(\rho,\sigma)=h(\epsilon)$, so that (\ref{disc2}) saturates the bound (\ref{qTUR}). In general, however, identity (\ref{disc4}) does not hold, not even for two level systems, as we show in the Monte Carlo simulations below.  

{\bf \emph{Simulations -}} Motivated by the minimal system that saturates the bound (\ref{disc4}), we now test numerically our main result (\ref{qTUR}) for two qubits $\rho$, $\sigma$ including quantum coherence. For each run, we draw a random operators ($\rho, \sigma, \hat{\theta}$). Then, we compute $U=U(\hat{\theta};\rho,\sigma)$ as in (\ref{disc0}) and $\tilde{S}=\tilde{S}(\rho,\sigma)$. 

For the simulation, we denote $X \sim I_x$ 
a random variable uniformly distributed in the interval $I_x$. We consider the decomposition $\rho = (1-p_1)
|0\rangle \langle 0| + p_1 |1\rangle \langle 1|$, where $p_1 \sim [0,1]$ for each run. Similarly, for each run, we independently draw a random $\sigma = (1-q_1)
|0\rangle \langle 0| + q_1 |1\rangle \langle 1| + C|0\rangle \langle 1| + C^*|1\rangle \langle 0|$, where $q_1 \sim [0,1]$, with $C:=|C|\exp(\phi_1 i)$, where $|C|^2 \sim [0,q_1(1-q_1)]$, $\phi_1 \sim [0,2\pi)$, so that $\sigma$ is completely positive. Finally, we draw a random Hermitian operator $\hat{\theta}=\omega(|1\rangle \langle 1| - |0\rangle \langle 0|) + D|0\rangle \langle 1| + D^*|1\rangle \langle 0|$, where $\omega \sim [0,1]$ and $D:=|D|\exp(\phi_2 i)$, with $|D|^2 \sim [0,1]$, $\phi_2 \sim [0,2\pi)$. Then, for each run, we plot a pair $(U,\tilde{S})$ as a single blue point in Fig.~1 and repeat the process for $10^4$ runs. One can see that our main result (\ref{qTUR}) is validated, $U\geq f(\tilde{S})$ for all runs. Some of them touch the bound, as expected, since the minimal system described in (\ref{disc4}) can be randomly drawn in this setup.

We also check the role of coherence between $\rho$ and $\sigma$ in our main result (\ref{qTUR}). In this case, we start by splitting $\tilde{S}(\rho,\sigma)$ in two positive contributions, 
\cite{Baumgratz2014,Streltsov2016a}
\begin{equation}
\label{sim1}
S(\rho||\sigma)=S(\Delta_\sigma \rho || \sigma) + C_\sigma (\rho),
\end{equation}

\begin{figure}[htp]
\includegraphics[width=3.3 in]{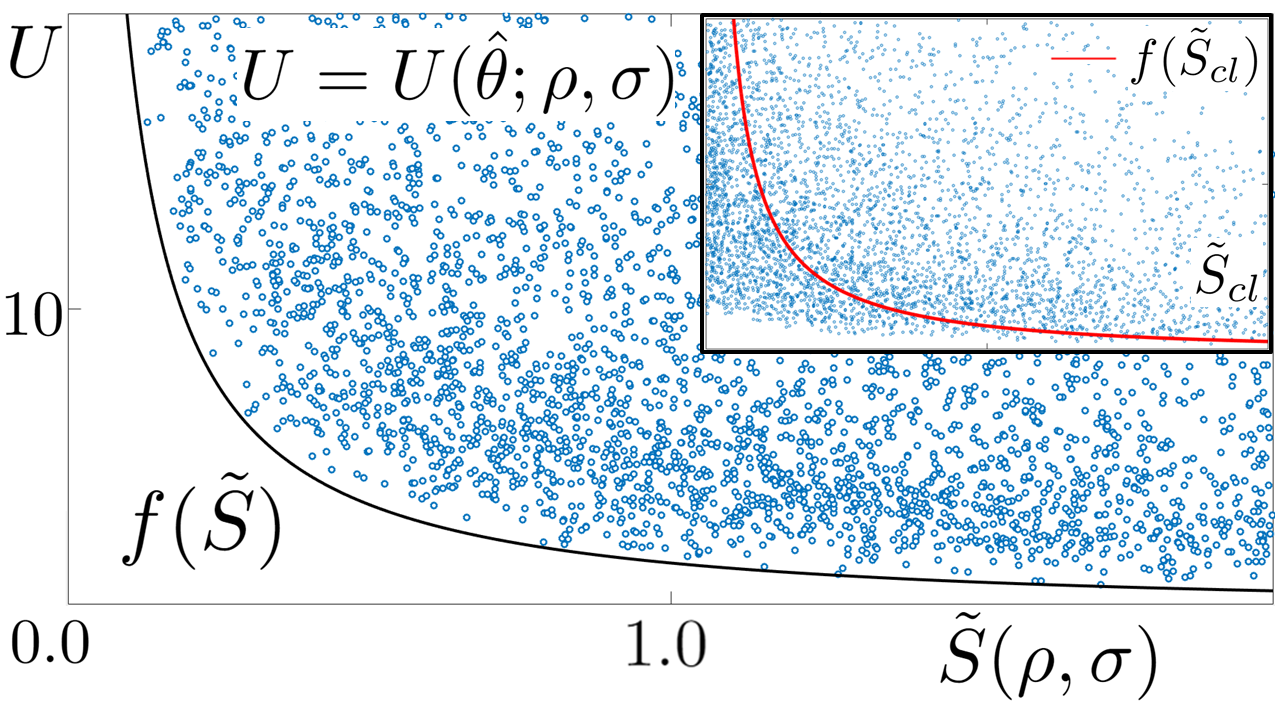}
\caption{(Color online) Monte Carlo simulation of the uncertainty $U=U(\hat{\theta};\rho,\sigma)$ as a function of the symmetric quantum relative entropy $\tilde{S}(\rho,\sigma)=[S(\rho||\sigma)+S(\sigma||\rho)]/2$. Each one of the $n=10^4$ blue points is a pair $(U,\tilde{S})$ computed for the random qubits $\rho, \sigma$ and random Hermitian operator $\hat{\theta}$. The lower bound $f(\tilde{S})$ from (\ref{qTUR}) is depicted in the solid black line, confirming $U\geq f(\tilde{S})$. The inset shows the same uncertainty $U$ vs. $\tilde{S}_{cl}$, which represents the classic component $\tilde{S}$ that disregards coherence between $\rho$ and $\sigma$, where the uncertainty clearly violates the classic bound $f(\tilde{S}_{cl})$ in solid red.}
\label{fig1}
\end{figure}
where $C_\sigma(\rho)=S(\Delta_\sigma(\rho))-S(\rho)$ is the relative entropy of coherence, $S(\rho)=-\tr(\rho \log \rho)$ is the entropy,  $\Delta_\sigma (\rho) := \sum_j |q_j\rangle \langle q_j| (\langle q_j|\rho|q_j\rangle)$ is a dephasing map in the basis of $\sigma$. In this case, we define $\tilde{S}_{cl}(\rho,\sigma)=[S(\Delta_\sigma \rho||\sigma) )+S(\Delta_\rho \sigma||\rho)]/2$
\begin{equation}
\label{sim2}
\tilde{S}(\rho,\sigma)=\tilde{S}_{cl}(\rho,\sigma) + \frac{1}{2}(C_\rho (\sigma) +C_\sigma(\rho)),
\end{equation}
where the absence of coherence between $\rho$ and $\sigma$, $[\rho,\sigma]=0$, makes $\tilde{S}(\rho,\sigma)=\tilde{S}_{cl}(\rho,\sigma)$. In the general case, one has
\begin{equation}
\label{sim3}
\tilde{S}(\rho,\sigma)\geq \tilde{S}_{cl}(\rho,\sigma) \rightarrow f(\tilde{S}(\rho,\sigma)) \leq f(\tilde{S}_{cl}(\rho,\sigma)),
\end{equation}
because $f$ is decreasing. Note that we have both $U(\hat{\theta};\rho,\sigma)\geq f(\tilde{S}(\rho,\sigma))$ from (\ref{qTUR}) and $f(\tilde{S}_{cl}(\rho,\sigma)) \geq f(\tilde{S}(\rho,\sigma))$ from (\ref{sim3}), so it is tempting 
to check if $f(\tilde{S}_{cl})$ is a viable (and possibly more efficient) lower bound for $U(\hat{\theta};\rho,\theta)$. If this is the case, then the coherence between $\rho$ and $\sigma$ could be ignored in the uncertainty relation, as we could just use $\tilde{S}_{cl}$ instead of $\tilde{S}$. To check this, the inset of Fig.~1 shows $U(\hat{\theta};\rho,\sigma)$ vs. $\tilde{S}_{cl}$, where $f(\tilde{S}_{cl})$ is depicted in solid red. For several runs, one can see that $U\geq f(\tilde{S}_{cl})$ is not true, where in all of them we have $U\geq f(\tilde{S})$, showing that we need to take coherence between $\rho$ and $\sigma$ into account for the uncertainty relation (\ref{qTUR}) to hold.

{\bf \emph{Application - quantum channels }}
An interesting application of (\ref{qTUR}) is obtained considering a completely positive trace preserving (CPTP) map $\mathcal{E}_t$. In this case, we have from the data processing inequality $\tilde{S}(\mathcal{E}_t(\rho),\mathcal{E}_t(\sigma)) \leq \tilde{S}(\rho,\sigma)$. Using that $f$ is decreasing, we have $ f(\tilde{S}(\mathcal{E}_t(\rho),\mathcal{E}_t(\sigma))) \geq f(\tilde{S}(\rho,\sigma))$. In this case, the bound (\ref{qTUR}) has a looser form in terms of initial conditions,
\begin{equation}
\label{channel1}
U(\hat{\theta};\rho(t),\sigma(t)) \geq f\big(\tilde{S}(\rho(t),\sigma(t))\big) \geq f\big(\tilde{S}(\rho(0),\sigma(0))\big),
\end{equation}
for any CPTP map $\mathcal{E}_t$, where $\rho(t)=\mathcal{E}_t(\rho(0))$ and $\sigma(t)=\mathcal{E}_t(\sigma(0))$ and $t \geq 0$ is a time parameter. The time dependent statistics of any observable $\hat{\theta}$ has a lower bound that depends on initial conditions only, but not on the dynamics $\mathcal{E}_t$. Particularly, if $\rho^*$ is a fixed point of the dynamics $\mathcal{E}_t$, we have $\mathcal{E}_t(\rho^*)=\rho^*$, then using (\ref{channel1}) with $\sigma(0)=\rho^*$ results in
\begin{equation}
\label{channel2}
U(\hat{\theta};\rho(t),\rho^*) \geq f\big(\tilde{S}(\rho(0),\rho^*)\big),
\end{equation}
in which the bound is also a constant in time as it depends solely on the dissimilarity between the initial state $\rho(0)$ and the fixed point $\rho^*$.

{\bf \emph{Application - quantum thermodynamics}} Also note that the specific choice $\hat{\theta}\rightarrow \log\rho_E$, $\rho\rightarrow \rho_E$ and $\sigma\rightarrow \rho_E'=\tr_S(\mathcal{U}(\rho_S \otimes \rho_E)\mathcal{U^\dagger})$ in the main result (\ref{qTUR}) yields
\begin{equation}
\label{Onsagerslike0}
\frac{\chi + \chi'}{(1/2)\Phi^2} \geq f\big(\frac{S(\rho_E'||\rho_E)+S(\rho_E||\rho_E')}{2}\big) \geq f(\frac{\Sigma+\Sigma^*}{2}),
\end{equation}
where $\Phi:=\tr_E((\rho_E -\rho_E')\log \rho_E)$ is the entropy flux \cite{RevModPhys.93.035008}, with generalized capacities $\chi:=\langle \log \rho_E^2 \rangle_{\rho_E} - \langle \log \rho_E \rangle_{\rho_E}^2$, $\chi':=\langle \log \rho_E^2 \rangle_{\rho_E'} - \langle \log \rho_E \rangle_{\rho_E'}^2$, and the last inequality comes from $\Sigma + \Sigma^* \geq S(\rho_E'||\rho_E) + S(\rho_E||\rho_E')$ and $f$ is decreasing. Using the inversion (\ref{disc1}) in (\ref{Onsagerslike0}), one also gets
\begin{equation}
\label{Onsagerslike1}
\frac{\Sigma+\Sigma^*}{2} \geq B\Big(\frac{2(\chi + \chi')}{\Phi^2}\Big),
\end{equation}
which is a general relation in quantum thermodynamics involving the entropy production and flux.

Now we briefly discuss the physical interpretation of $\Sigma^*$. We consider the quantum trajectory of four measurements, following the stochastic treatment of \cite{RevModPhys.93.035008,Manzano2018PRX}. In this case, $\gamma=\{m,\nu',n,\nu\}$, where ($m,\nu'$) represents the outcomes of the initial measurement in the basis $|\psi_m \rangle \otimes |\nu' \rangle$, built from the eigenbasis of $\rho_S' = \sum_m p_m' |\psi_m \rangle \langle \psi_m|$ and $\rho_E= \sum_\nu q_{\nu} |\nu \rangle \langle \nu |$. The pair ($n,\nu$) represents the final measurement in the basis $|n \rangle \otimes |\nu \rangle$, built from the eigenbasis of $\rho_S = \sum_n p_n |n\rangle \langle n |$ and $\rho_E$. Note that both initial and final local measurements of he environment are performed in the same basis. 

Now we take the initial state as $\rho_S' \otimes \rho_E$, perform the first measurement, yielding ($m,\nu'$), apply the unitary $\mathcal{U}^\dagger$ and perform the second measurement, yielding ($n,\nu$), such that the forward probability is defined as $P_F(\gamma)=|\langle n,\nu | \mathcal{U^\dagger}|\psi_m,\nu'\rangle|^2 p_m' q_{\nu'}$. Now for the backward process, we consider the initial state $\tilde{\rho}:=\rho_S \otimes \rho_E$, perform the first measurement with value $(n,\nu)$, then the unitary $\mathcal{U}$ and the final measurement, $(m,\nu')$, which results in the probability of the backward process, $P_B(\gamma)=|\langle  \psi_m,\nu' | \mathcal{U} |n,\nu\rangle|^2 \tilde{\rho}_{n\nu}$, where $\tilde{\rho}_{m \nu}:=\langle n,\nu | \tilde{\rho} |n,\nu \rangle = p_n q_\nu$. Finally, we define the average stochastic entropy production for the path probabilities $P_F(\gamma)$ and $P_B(\gamma)$, $\langle \sigma \rangle := D(P_F|P_B)$, resulting in
\begin{equation}
\langle \sigma \rangle = \sum |\langle n,\nu | \mathcal{U^\dagger}|\psi_m,\nu'\rangle|^2 p_m'q_{\nu'} \ln (\frac{p_m' q_{\nu'}}{p_n q_\nu}) = \Sigma^*,
\end{equation}
after some manipulation, using (\ref{qthermo2}). Note that we used the same measurement scheme for the reservoir in both ends of the path, as suggested in the original derivation of $\Sigma$ \cite{Manzano2018PRX,RevModPhys.93.035008}. However, in the derivation of $\Sigma$, it is used a different initial state for the backward process. For that reason, although $\Sigma^*$ has a stochastic interpretation, it relies on a specific choice of backward process that differs from the original protocol for the definition of $\Sigma$. Thus, $\Sigma^*$ is not a entropy production in the sense of (\ref{qthermo1}) in the general case.

{\bf \emph{Conclusions - }}
We have proposed an uncertainty relation on quantum information (\ref{qTUR}). The theorem states that a certain statistics of any Hermitian operator $\hat{\theta}$ has a lower bound in terms of the quantum relative entropies between $\rho$ and $\sigma$. We verified the bound for Monte Carlo simulations using two random qubits and random operators in the presence of coherence, where the saturation of the bound and the role of coherence was discussed. We also applied the result for general quantum channels, obtaining a lower bound for the time dependent uncertainty in terms of the initial conditions (\ref{channel1}), and the fixed point (\ref{channel2}). Finally, we applied the result in the most general setup of quantum thermodynamics, obtaining a quantum thermodynamic uncertainty relation in terms of the quantum entropy production and its dual (\ref{qthermo3}).

{\bf \emph{Appendix - }} We used an expression (\ref{form10}) from information theory that connects observables and divergences in the form of a uncertainty relation. The original idea \cite{Salazar2022d,Nishiyama2022} uses real observables and here we need to fix it for complex ones, although the proof is essentially the same of \cite{Salazar2022d}.
Consider probabilities $P,Q$ in $s \in S$, $\sum_s P(s)= \sum_s Q(s)=1$ and a complex valued random variable $\theta(s) \in \mathbb{C}$. If $P$ and $Q$ are not absolute continuous ($P(s)>0$ and $Q(s)=0$ or $P(s)=0$ and $Q(s)>0$ for some $s\in S$), then (\ref{form10}) is trivial, because $D(P|Q)+D(Q|P)=\infty$ and $f(\infty)=0$. So we consider the relevant case which is $P(s)=0 \iff Q(s)=0$. We define $S'=\{s \in S| P(s)+Q(s)>0\}$ and the probability $\tilde{P}(s):=(P(s)+Q(s))/2$ in $S'$, $\sum_{s \in S'}\tilde{P}(s)=1$, $\overline{\theta}_X:=\langle \theta \rangle_X=\sum_s \theta(s) X(s)$, for $X \in \{P,Q,\tilde{P}\}$. 
Note that the expression $|\overline{\theta}_P - \overline{\theta}_Q|^2$ can be rewritten as
\begin{equation}
\label{app1}
\frac{1}{4}|\overline{\theta}_P - \overline{\theta}_Q|^2 = |\sum_{s\in S'} (\theta(s)-c) \frac{(P(s)-Q(s))}{2}|^2,
\end{equation}
for any complex $c$. Using Cauchy-Schwarz inequality, we also obtain for any complex $c$,
\begin{equation}
\label{app2}
|\sum_{s\in S'} (\theta(s)-c) \frac{(P(s)-Q(s))}{2}|^2 \leq \langle (|\theta-c|^2 \rangle_{\tilde{P}}\langle (\frac{P-Q}{P+Q})^2\rangle_{\tilde{P}},
\end{equation}
so that combining (\ref{app1}) and (\ref{app2}) for $c=\overline{\theta}_{\tilde{P}}$, it yields 
\begin{equation}
\label{app3}
\frac{1}{4}|\overline{\theta}_P - \overline{\theta}_Q|^2 \leq \langle (|\theta-\overline{\theta}_{\tilde{P}}|^2 \rangle_{\tilde{P}}\langle (\frac{P-Q}{P+Q})^2\rangle_{\tilde{P}}.
\end{equation}
Finally, as showed in \cite{Salazar2022d}, we use the results
\begin{equation}
\label{app4}
\langle (\frac{P-Q}{P+Q})^2\rangle_{\tilde{P}} \leq \tanh^2[(1/2)g(\tilde{D}(P,Q))],
\end{equation}
where $\tilde{D}(P,Q)=[(D(P|Q)+D(Q|P)]/2$, and the identity
\begin{equation}
\label{app5}
4\langle |\theta-\overline{\theta}_{\tilde{P}}|^2 \rangle_{\tilde{P}} = 2(\langle|\theta|^2\rangle_P-|\overline{\theta}_P|^2) + 2(\langle|\theta|^2\rangle_Q-|\overline{\theta}_Q|^2) + |\overline{\theta}_P - \overline{\theta}_Q|^2. 
\end{equation}
Combining (\ref{app3}), (\ref{app4}) and (\ref{app5}) it results in (\ref{form10}).

\bibliography{lib7}
\end{document}